\documentclass[footinbib,eqsecnum,twocolumn,showpacs,amsmath,amssymb]{revtex4}
\usepackage{natbib}
\usepackage{bm}
\usepackage{graphicx}
\usepackage{amsfonts}
\begin{document}
\title{Budding and Domain Shape Transformations in \\
  Mixed Lipid Films and Bilayer Membranes} \author{J.~L.~Harden}
\email{harden@jhu.edu} \affiliation{Department of Chemical
  Engineering, Johns Hopkins University Baltimore MD 21218-2689, USA}
\author{F.~C.~MacKintosh} \email{fcm@nat.vu.nl} \affiliation{Division
  of Physics \& Astronomy, Vrije Universiteit, 1081 HV Amsterdam, The
  Netherlands} \author{P.~D.~Olmsted} \email{p.d.olmsted@leeds.ac.uk}
\affiliation{School of Physics \& Astronomy and Polymer IRC,
  University of Leeds, Leeds LS2 9JT, UK} \date{\today}
\begin{abstract}
  We study the stability and shapes of domains with spontaneous
  curvature in fluid films and membranes, embedded in a surrounding
  membrane with zero spontaneous curvature.  These domains can result
  from the inclusion of an impurity in a fluid membrane, or from phase
  separation within the membrane.  We show that for small but finite
  line and surface tensions and for finite spontaneous curvatures, an
  equilibrium phase of protruding circular domains is obtained at low
  impurity concentrations.  At higher concentrations, we predict a
  transition from circular domains, or \emph{caplets}, to stripes.  In both
  cases, we calculate the shapes of these domains within the Monge
  representation for the membrane shape.  With increasing line
  tension, we show numerically that there is a budding transformation
  from stable protruding circular domains to spherical buds. We
  calculate the full phase diagram, and demonstrate a two triple
  points, of respectively bud-flat-caplet and flat-stripe-caplet
  coexistence.
\end{abstract}
\pacs{{87.16.Dg}{Membranes, bilayers, and vesicles}}
\pacs{{68.05.-n} {Liquid-liquid interfaces}}
\pacs{{68.18.Jk} {Phase transitions}}
\maketitle
%%%%%%%%%%%%%%%%%%%%%%%%%%%%%%%%%%%%%%%%%%%%%%%%%%%%%%%%%%%%%%%%%%%%%%%%%
\section{Introduction}
%%%%%%%%%%%%%%%%%%%%%%%%%%%%%%%%%%%%%%%%%%%%%%%%%%%%%%%%%%%%%%%%%%%%%%%
%%%%%%%%%%%%%%%%%%%%%%%%%%%%%%%%%%%%%%%%%%%%%%%%%%%%%%%%%%%%%%%%%%%%%%%

Fluid membranes occur in a wide variety of physical, chemical, and
biological systems
\cite{amphiphiles,bilayers,biomembranes,lipo-mann95}.  Examples
include surfactant films, uni-lamellar and multi-lamellar vesicles,
and lipid bilayer membranes, such as those in biological cells.
Increased attention has been paid to the properties of multi-component
bilayer membranes.  There are at least two important reasons for this.
On the one hand, biological cell membranes naturally involve mixtures
of several different lipid and protein components
\cite{biomembranes,lipo-mann95}.
Processes such as budding, shape and textural transformations, and
raft formation are believed to involve local inhomogeneities in these
mixtures
\cite{farge92,Dobe+93,SackFede95,mcconnell99a,farge00,crowe01,webb03,keller02,keller03,keller04,munro03,edidin_rev03,gaus03a}.
On the other hand, lipid mixtures have been shown to play an important
role in the formation of stable vesicles for a variety of potential
biomedical applications such as controlled gene and drug delivery
\cite{lipogenes97,lipomedicine98}.

The additional internal degrees of freedom that accompany the presence
of two or more components within a fluid membrane can lead to a rich
set of different membrane properties.  The understanding of the
effects of composition variations on membrane properties is important,
as it may shed light on the behavior of cell membranes, and also
enable rational control of synthetic membrane structure and function.
The compositional degrees of freedom in a multi-component membrane can
dramatically influence both its morphology and phase behavior.  For
instance, phase separation within the membrane can occur, analogous to
phase separation of ordinary fluids.  However, such two-dimensional
phase separation within a membrane is expected to be closely linked
with the three-dimensional shape of the membrane.  This is because, at
a microscopic level, the curvature properties, and especially the
spontaneous curvature, are largely dictated by the structure of the
constituent molecules~\cite{israelachvili}.  Thus, regions of
different composition will often have different curvature properties.
In such cases, the formation of a domain of one phase within a matrix
of another phase can lead to a localized deformation of the membrane.
This deformation can be enhanced by finite line tension between
domains, which when sufficiently large can drive budding (the
formation and subsequent separation of a small vesicle).  Experimental
studies of model two and three component systems have examined domain
formation~\cite{farge92,SackFede95,farge00,mcconnell99a,crowe01,webb03,keller02,keller03,keller04}
and its role in the process of membrane shape
transformations~\cite{webb03,keller03} and
budding~\cite{farge92,SackFede95}.

Many theoretical studies have examined the interplay between internal
degrees of freedom and membrane shape in two-component fluid
membranes~\cite{leibler86,leibler87,safranpinc90,safranpinc91,AKK92,Lipo92b,MackSafr93,Seif93,JuliLipo93,Kawa+93,Tani+94,
HardMack94,JuliLipo96, kumar99,gozdz01}.  In theoretical studies of
phase-separated bilayer vesicles, it was found that budding can occur
in the limit of large line tension between the two
phases~\cite{Lipo92b,JuliLipo93,JuliLipo96}.  However, when surface
tension is relevant, phase separation can lead to stable, modulated
phases of flat films and
vesicles 
\cite{leibler87,AKK92,Kawa+93,Tani+94,HardMack94,JuliLipo96,kumar99,gozdz01}.
Indeed, equilibrium modulated morphologies have been observed in
several recent studies of phase-separated multi-component bilayer
vesicles~\cite{webb03,keller03}.  On the other hand, the deformation
of homogeneously mixed two-component fluid membranes may induce
in-plane phase separation\cite{Seif93}.

\begin{figure}[ht]
\displaywidth\columnwidth
\includegraphics[scale=0.35]{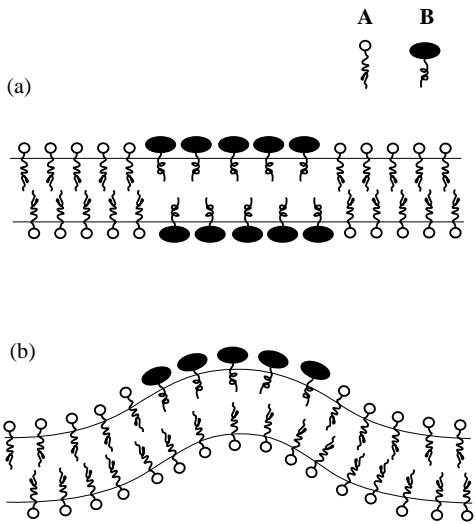}
\caption{
(a) Flat symmetric phase-separated membrane, and (b) curved,
asymmetric phase-separated membrane.}
\label{pics}
\end{figure}

For a mixed monolayer, the deformation of the film due to phase
separation of two components with differing spontaneous curvature is
an immediate consequence of the different molecular architectures of
the two components.  For a bilayer membrane, however, the coupling of
composition to curvature is somewhat more subtle because of symmetry
considerations.  Phase separation can take two different forms within
a bilayer membrane.
As shown in Fig.~\ref{pics}(a), a symmetric domain of one component
can develop within a matrix of another phase.  In Fig.~\ref{pics}(b),
an asymmetric domain is shown, in which the two halves of the bilayer
have different compositions.  As illustrated in the figure, these
effects can influence the shapes and phase behavior of membranes in
different ways.  In the symmetric case [Fig.~\ref{pics}(a)], the
domain can remain flat \cite{Lipo92b} although the domain can be more
or less rigid than the surrounding membrane, a situation reminiscent
of rafts in biological membranes.  However, composition
inhomogeneities can give rise to deformation of a bilayer membrane if
there is an asymmetric distribution of the lipid constituents across
the bilayer, as shown in Fig.~\ref{pics}(b) \cite{gebhardt77}.  Such
lipid asymmetry has been clearly demonstrated experimentally in
biological membranes \cite{opdenkamp79,verklejj}.  This broken
symmetry of the bilayer can either arise spontaneously
\cite{MackSafr93} or as a result of different environments on the
inside and outside of a vesicle.  In the latter case, we must also
consider the possibility of a non-zero surface tension due to an
osmotic pressure difference between the inside and outside.

%%%%%%%%%%%%%%%%%%%%%%%%%%%%%%%%%%%%%%%%%
%%%%%%%%%%%%%%%%%%%%%%%%%%%%%%%%%%%%%%%%%%

\begin{figure}
\includegraphics[width=2.8truein]{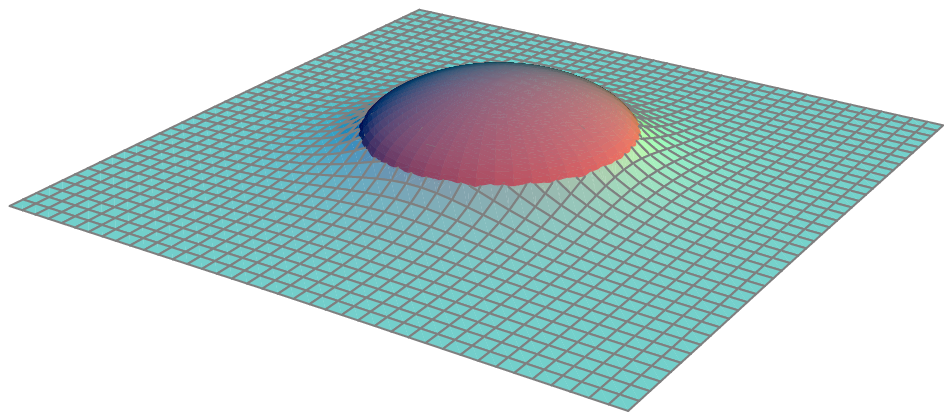}
\includegraphics[width=2.8truein]{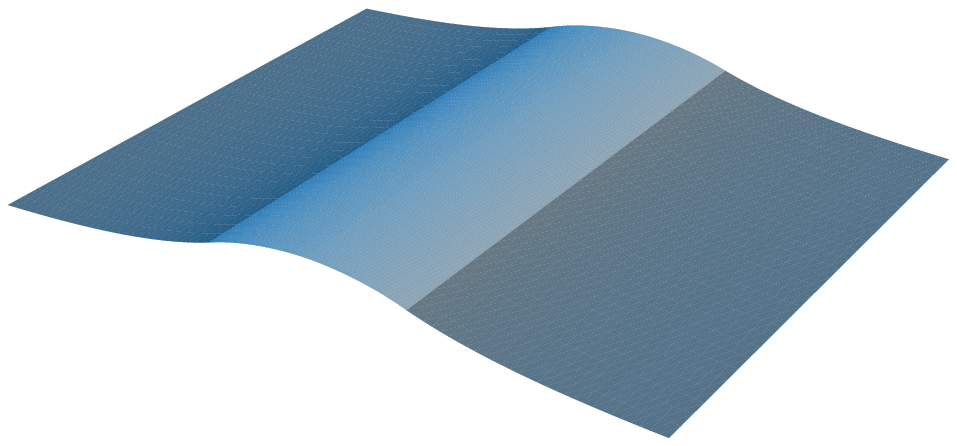}
\includegraphics[width=2.8truein]{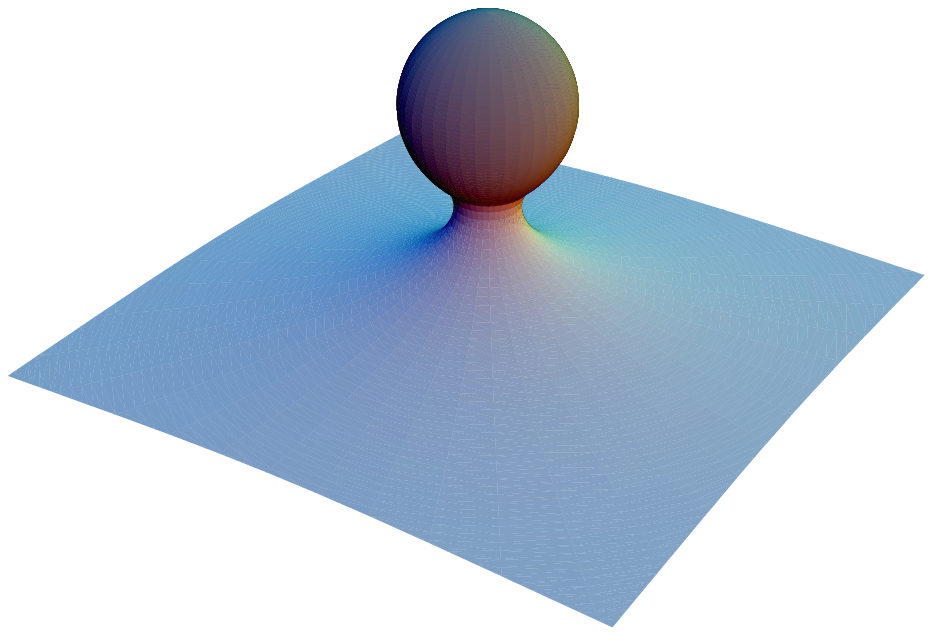}
\caption{
  Schematic representation of phase-separated circular (top) and
  stripe (middle) domains and a bud (bottom) of the $B$ component,
  with induced deformations.}
\label{shapes}
\end{figure}

In this paper, we examine the stability and equilibrium shapes of
domains in asymptotically flat fluid films (or giant vesicles with
dimensions much larger than the domain sizes).  We focus on domains of
constituents with finite effective spontaneous curvature, embedded in
a matrix of membrane material with zero spontaneous curvature.  These
domains can either be due to phase separation of two or more
components in a mixed membrane, or to the inclusion of an ``impurity''
in the film (such as a membrane protein or a surface adsorbed
macromolecule) \cite{Sens2004}.  We show that for membranes under
tension, stable protruding circular domains (we call these \emph{caplets}, as in Refs.\ \cite{gozdz01,HardMack94}) similar to
those observed for mixtures of lecithin and phosphatitic acid
\cite{gebhardt77} can occur at low concentrations of the minority
component or impurity.  We illustrate the structure of such circular
\emph{caplet} domains in Fig.~\ref{shapes}.
%%%%%%%%%%%%%%%%%%%%%%%%%%%%%%%%%%%%%%%%%%%%%%%%%%%%%%%%%%%%%%%%%%
%%%%%%%%%%%%%%%%%%%%%%%%%%%%%%%%%%%%%%%%%%%%%%%%%%%%%%%%%%%%%%%%%%
We also demonstrate the possibility of a phase transition from a
mesophase of circular caplets to a mesophase of protruding stripe
domains either with increasing area fraction of the minority component
or with increasing tension.  We illustrate these
stripe domains in Fig.~\ref{shapes}.

In both cases, the domains are shown to be stable with respect to
complete phase separation in a flat membrane.  In
Section~\ref{sec:model} we present our model and briefly discuss our
analysis in the Monge gauge treatment. In Section~\ref{sec:eqns}, we
provide the resulting Euler-Lagrange equations and boundary conditions
used in our Monge gauge calculations.  In Section~\ref{sec:regimes} we
present Monge limit results in the dilute and concentrated regimes,
and discuss the extent to which the limiting approximations of Ref.
\cite{HardMack94} apply. In Section~\ref{sec:num} we present the
results of a numerical calculation of the budding transition beyond
the Monge approximation, and we conclude in Section~\ref{sec:yadayada}
with a summary and discussion.

%%%%%%%%%%%%%%%%%%%%%%%%%%%%%%%%%%%%%%%%%%%%%%%%%%%%%%%%%%%%%%%%%%%%%%%
\section{Model (Monge Gauge)}\label{sec:model}
%%%%%%%%%%%%%%%%%%%%%%%%%%%%%%%%%%%%%%%%%%%%%%%%%%%%%%%%%%%%%%%%%%%%%%%

We consider a single idealized two-component membrane consisting of
incompatible amphiphiles $A$ and $B$.  Furthermore, we consider asymmetric
bilayers, in which phase separation occurs in only one leaflet of the
bilayer, as shown in Fig.~\ref{pics}(b).  This applies, for instance,
to the case of phase separation in an asymmetric bilayer vesicle or
biomembrane.  Such domains can also be induced by the adsorption of
macroions on the inside or outside of a bilayer vesicle, as in the
experiments of Refs.~\cite{Baek+95,SackFede95}.  We assume that the
binary liquid mixture is far from its critical point, implying that
there are sharp interfaces between the domains of $A$ and $B$, and that
the membrane consists of a fluid minority phase $B$ surrounded by an
asymptotically flat fluid majority phase $A$.

\begin{figure}[ht]
\includegraphics[scale=1.0]{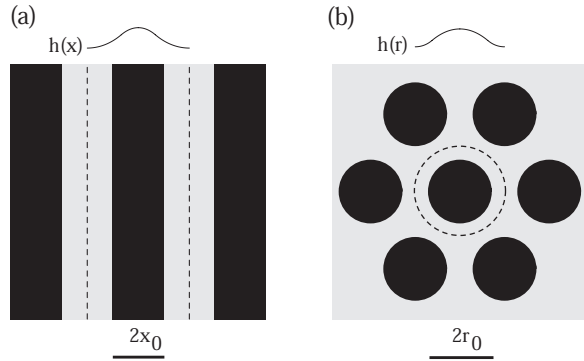}
\caption{
  A sketch of the stripe and caplet morphologies.  Fig.~(a) shows a
  linear array of ${\rm B}$-phase stripes of width $2 x_0$ (black) in a
  matrix of phase ${\rm A}$ (grey).  Fig.~(b) shows a hexagonal array
  of ${\rm B}$-phase circular caplets of radius $r_0$ (black) in a
  matrix of phase ${\rm A}$ (grey).  The dashed lines in each figure
  show the appropriate Wigner-Seitz cell boundaries, while the
  membrane displacement profile of an appropriate unit cell cross
  section is shown above each figure.}
\label{domains}
\end{figure}

We investigate the properties of two prototypical domain morphologies
(see Fig.~\ref{domains}), which are idealizations of actual
structures that may exhibit disordered textures of polydisperse,
irregular domains:
\begin{itemize}
\item[i)] Quasi-one-dimensional domains of ${\rm B}$ (stripes),
  with an impurity phase of width $2 x_0$ embedded inside a majority
  stripe with total with $2X_0$. The stripe length (system dimension)
  is $L$, and the projected area of an element (minority plus majority
  repeat unit) is $S_{A,\perp}+S_{B,\perp}=2 X_0L$.
\item[ii)] Monodisperse circular domains of ${\rm B}$ (caplets).
  For caplets we take a circular Wigner-Seitz cells with impurity
  radius $r_0$ and total radius $R_0$.  The circular Wigner-Seitz cell
  approximation leads to a lower bound for the free energy of a less
  symmetric structure, such as a hexagonal array.  The projected area
  of an element is $S_{A,\perp}+S_{B\,\perp}=\pi R_0^2$.
\end{itemize}
Below we will typically scale these dimensions by the spontaneous
curvature $c_0$, thus defining $\rho_0=r_0c_0$, $\bar{R}_0=R_0c_0$, $\xi_0=x_0c_0$, $\Xi_0=X_0c_0$.

%%%%%%%%%%%%%%%%%%%%%%%%%%%%%%%%%%%%%%%%%%%%%%%%%%%%%%%%%%%%%%%%%%%%%%%
%%%%%%%%%%%%%%%%%%%%%%%%%%%%%%%%%%%%%%%%%%%%%%%%%%%%%%%%%%%%%%%%%%%%%%%
For given material parameters, we obtain minimum energy height
profiles $h({\rm\bf r})$ (within a linearized Monge gauge
description), where ${\rm\bf r}$ is the position in the plane, and
minimize the resulting free energy over the domain dimensions. We
consider the background ${\rm A}$ phase to be flat and restrict our
attention to the simplified case of equal bending moduli $\kappa_{{\rm
    A}}=\kappa_{\rm B} \equiv\kappa$.

For phase separated membranes under tension,
the free energy of a single domain (stripe or caplet)  has the form
\begin{equation}
G_0=F_l + F_s + F_b
\label{Eq1}
\end{equation}
where $F_l$, $F_{s}$, and $F_b$ are the energy contributions from the
interfacial (line) tension between the two phases, the membrane frame
(surface) tension, and the bending elasticity.  To lowest order in the
Monge representation (see Appendix \ref{app:a}), valid for nearly flat
membranes, these contributions relative to a flat reference state are
given as follows.
\begin{widetext}
The bending energy is
\begin{equation}
F_b=\frac12\kappa\left\{
\int_{S_{\rm A}} \left(\nabla^2 h\right)^2 dS_{\perp}
+\int_{S_{\rm B}} \left[ \left(\nabla^2 h\right)^2
-2c_0\nabla^2 h + c_0^2\left(\nabla h\right)^2\right] dS_{\perp}
\right\},
\label{Eq4}
\end{equation}
where $dS_{\perp}$ is the (flat) area measure in the plane, $S_i$
denotes the projected area occupied by phase $i$, $h({\rm\bf r})$ is
the height profile above some flat reference plane, $\nabla$ is a
two-dimensional gradient operator, $\kappa$ is the bending modulus,
and $c_0$ is the spontaneous curvature of the ${\rm B}$ phase.  The
line tension energy is
\begin{align}
F_l &= \lambda \oint_{\partial S_{\rm B}}dl =
\left\{
\begin{array}{ll}
\lambda\, 2\pi r_0 N& \qquad\hbox{(caplets)}\\
\noalign{\vskip5truept}
\lambda\, 2 L N& \qquad\hbox{(stripes)}\\
\end{array}
\right.
\label{Eq2}
\end{align}
where $N$ is the number of domains, $\partial S_{\rm B}$ denotes the
boundary of the impurity region, and $\lambda$ is the line tension.
The work done by the membrane in deforming relative to the flat state
is
\begin{equation}
F_{s}= \frac12\left\{
\sigma\int_{S_{\rm A}} \left|\boldsymbol{\nabla} h\right|^2 dS_{\perp}
+ (\sigma+\mu)\int_{S_{\rm B}} \left|\boldsymbol{\nabla}
  h \right|^2 
dS_{\perp}\right\},
\label{Eq3}
\end{equation}
\end{widetext}
where $\sigma$ is the frame tension, and the Lagrange multiplier (or
exchange chemical potential per unit area) $\mu$ controls the amount
of impurity phase. We assume a fixed area per molecule in the
membrane, and thus ignore stretching energies.

It will be convenient to define a renormalized chemical potential
$\tilde{\mu}$, by
\begin{align} \tilde{\mu}&=\mu + \kappa c_0^2, \label{eq:3}
\end{align}
where $\kappa c_0^2$ arose in Eq.~(\ref{eq:3}) from keeping all terms
to second order in the height field \cite{gozdz01}. This
renormalization, neglected in \cite{HardMack94}, is typically very
small: $\kappa c_0^2$ is of order $k_BT$ divided by a curvature radius
squared of order $(100\textrm{nm})^2$, while $\sigma\sim k_BT$ divided
by a length squared of order $\textrm{nm}^2$.  In fact, we will find
that this renormalization does not change the phase boundaries if only
one phase has a non-zero spontaneous curvature; see \cite{gozdz01} for
a different example.

Stationarity of Eq.~(\ref{Eq1}) with respect to height variations
leads to Euler-Lagrange (E-L) equations for the height profiles in
regions ${\rm A}$ and ${\rm B}$.  The detailed derivation of the these
equations are presented in Appendix~\ref{app:a}.  To calculate the phase
behavior we must  minimize the free energy $G$ of an array of domains,
\begin{equation}
G(\mu,\sigma,\lambda) = N
G_0(\mu,\sigma,\lambda), 
\end{equation}
where
\begin{equation}
N = \frac{S_{fr}}{S_{A,\perp}+S_{B\,\perp}}
\end{equation}
is the number of domains and $S_{fr}$ is the frame area, which without
loss of generality we consider to be fixed. The free energy must also
be minimized over the dimensions of the ${\rm A}$ and ${\rm B}$
domains, $x_0$ and $X_0$ for the stripes and $r_0$ and $R_0$ for the
caplets:
\begin{subequations}
\label{ELequations}
  \begin{align}
    \frac{\partial G}{\partial x_0}&=0,&\
    \frac{\partial G}{\partial X_0}&=0,&\textrm{(stripes)}\label{ELS}\\
    \frac{\partial G}{\partial r_0}&=0,&\ 
    \frac{\partial G}{\partial R_0}&=0,&\textrm{(caplets)}.\label{ELC}
  \end{align}
\end{subequations}
Physically, minimization over the inner length ($x_0$ or $r_0$) is
equivalent to allowing exchange of $B$ for $A$ species, while minimization
over the outer length ($X_0$ or $R_0$) is equivalent to allowing more
or less total area into the system, doing work against the frame
tension.

Phase coexistence is found by equating  chemical potentials,
\begin{equation}
  \label{eq:1}
  \mu_{\rm stripes}=\mu_{\rm caplets}.
\end{equation}
In the dilute limit, where all the impurity is in a single domain, we
must minimize the free energy per impurity,
\begin{equation}
g(\mu,\sigma,\lambda) =
\frac{G_0(\mu,\sigma,\lambda)}{A^{\perp}_{\rm B}} 
\end{equation}
over the impurity domain size. In using the projected area of the
${\rm B}$ phase we have kept terms consistent with the Monge gauge
calculation (and hence ignored a term of order $h^2$ in the
denominator).

Following \cite{HardMack94}, we will use \emph{hatted} variables for
dimensionless quantities $\hat\lambda, \hat\sigma, \hat\mu$
(introduced below and Appendix~\ref{app:a}) scaled by $\kappa$ and
appropriate powers of $c_0$, and unadorned variables for physical
quantities. This differs from the more recent work of \cite{gozdz01}
in which dimensional quantities were hatted and dimensionless
quantities were not hatted.

For simplicity, we have also assumed that all bending moduli are the
same in the different phases.  Different mean curvature moduli
$\kappa$ will obviously shift the phase boundary in favor of the phase
with a lower curvature modulus, while a difference in Gaussian
curvature moduli would lead to a shape-dependent line tension that
will shift the phase boundaries \cite{JuliLipo96}.
%%%%%%%%%%%%%%%%%%%%%%%%%%%%%%%%%%%%%%%%%%%%%%%%%%%%%%%%%%%%%%%%%%%%%%%
\section{Euler-Lagrange equations and boundary conditions}\label{sec:eqns}
%%%%%%%%%%%%%%%%%%%%%%%%%%%%%%%%%%%%%%%%%%%%%%%%%%%%%%%%%%%%%%%%%%%%%%%
Variation of $G_0(\mu,\sigma,\lambda)$ with respect to $h({\rm\bf r})$
gives the Euler-Lagrange equations for the height profile. To
calculate the free energy we only need the slope $\eta=dh/dx$ (stripe)
or $\eta=dh/dr$ (caplet).  As shown in
Appendix~\ref{app:a}, the slopes in regions $A$ and $B$ ($i={\rm A},{\rm
  B}$) satisfy
\begin{subequations}
\begin{align}
\frac{d^2\eta}{d\xi^2}- {\zeta_i}^2\eta=0&\qquad\hbox{(stripes)}
\label{Eq5}\\
\frac{d^2\eta}{d\rho^2} + \frac{1}{\rho}\frac{d\eta}{d\rho} -
\left( {\zeta_i}^2+\frac{1}{\rho^2}\right)\eta=0 &
\qquad\hbox{(caplets)}
\label{Eq6}
\end{align}
\end{subequations}
where we have introduced dimensionless variables,
\begin{subequations}
\begin{eqnarray}
\xi &=& c_0 x \\
\rho &=& c_0 r\\
{\zeta}_{\rm B}^2 &=& \hat{\sigma}+\hat{\mu} \label{eq:24}\\
{\zeta}_{\rm A}^2 &=& \hat{\sigma} \\
\hat\sigma &=& \frac{\sigma}{\kappa c_0^2} \\
\hat{\mu} &=& \frac{\tilde{\mu}}{\kappa c_0^2} + 1 \\
{\hat{\lambda}} &=& \frac{\lambda}{\kappa c_0}.
\end{eqnarray}
\end{subequations}
Note that the renormalization of $\mu$ due to $\kappa c_0^2$ is
identical in both stripe and caplet phases, and will thus play no role
in determining coexistence, \textit{i.e.}, in Eq.~(\ref{eq:1}).

Symmetry of $\eta$ at the domain centers implies
\begin{equation}
\eta(0) = \eta''(0) = 0.
\label{eq:h2a}
\end{equation}
where primes denote derivatives with respect to either $\rho$ or
$\xi$. Another condition may be supplied by requiring a well-defined
energy.  A discontinuity in $\eta$ would imply singular curvature at
the boundary, leading to a non-physical curvature energy in the region
of the interface $\partial_{S_B}$ between $A$ and $B$ phases.  Hence
we must have
\begin{equation}
\left[\eta_{\rm A}- \eta_{\rm B}\right]_{\partial_{S_B}}=0.
\label{eq:h4a}
\end{equation}
The final boundary conditions follow from the variation of the height
profile, and are derived in Appendix~\ref{app:a}. In the dilute limit,
\begin{align}
\left[{c}_{\rm B} - {c}_{\rm A}\right]_{\partial_{S_B}} &= c_0
&&\textrm{(torque)}
\label{eq:b2}\\
\sigma \eta_{\rm A}(\infty) - \kappa {c}_{\rm A}'(\infty)  &= 0
&&\textrm{(normal force)}
\label{eq:b3}\\
{c}_{\rm A}(\infty) &= 0, \label{eq:b4}
&&\textrm{(torque)}
\end{align}
where $c(r)$ is the curvature:
\begin{equation}
  c = \left\{
    \begin{array}{c@{\qquad}l}
      \eta'&(\hbox{stripes})\\
      \noalign{\vskip6truept}
      {\displaystyle\frac{1}{\rho}\frac{\partial}
        {\partial\rho}\left(\rho\eta\right)}&(\hbox{caplets}).
    \end{array}\right.
\end{equation}
Eqs.~(\ref{eq:b2}) and~(\ref{eq:b4}) are torque balances at the ${\rm
  A}-{\rm B}$ interface \cite{KozlHelf95} and at the system boundary,
and Eqs.~(\ref{eq:b3}) are the consequence of zero vertical force at
the ${\rm A}-{\rm B}$ interface. For the concentrated limit, the last
two equations should be replaced by the equivalent result for periodic
boundary conditions:
\begin{equation}
\left.\eta_{\rm A}\right|_{\partial_{S_B}} =
\left.c'\right|_{\partial_{S_B}}=0\qquad\hbox{(concentrated regime)}.
\end{equation}

%%%%%%%%%%%%%%%%%%%%%%%%%%%%%%%%%%%%%%%%%%%%%%%%%%%%%%%%%%%%%%%%%%%%%%%
\section{Monge Limit: Results}\label{sec:regimes}
%%%%%%%%%%%%%%%%%%%%%%%%%%%%%%%%%%%%%%%%%%%%%%%%%%%%%%%%%%%%%%%%%%%%%%%
The general solutions to the Euler-Lagrange equations are given in
Appendix~\ref{app:b}. Before discussing the phase behavior in the
general (concentrated) case, we discuss the dilute limit for large
frame tensions, where several approximations simplify the results.
%%%%%%%%%%%%%%%%%%%%%%%%%%%%%%%%%%%%%%%%%%%%%%%%%%%%%%%%%%%%%%%%%%%%%%%
\subsection{Dilute Limit ($\hat{\sigma}\gg\hat{\mu}\simeq 0$)}
%%%%%%%%%%%%%%%%%%%%%%%%%%%%%%%%%%%%%%%%%%%%%%%%%%%%%%%%%%%%%%%%%%%%%%%

Since the height profiles decay in the ${\rm A}$ phase with a
characteristic length $\sqrt{\kappa/\sigma}$, the dilute limit is
valid provided that the domains are separated by more than this
length. The general solutions in the Monge gauge,
Eqs. ~(\ref{eq:sol1d}, \ref{eq:sol2d}), have inverse penetration depths
$\zeta_{\rm A}=\sqrt{\hat\sigma}$ and $\zeta_{\rm B}=\sqrt{\hat\sigma+\hat\mu}$. In the
dilute limit and for large frame tension we may ignore $\hat\mu$ relative
to $\hat{\sigma}$ in the penetration depths. 
Taking the limits
$R_0\rightarrow\infty$ and $X_0\rightarrow\infty$, the solutions (given in Appendix~\ref{app:b})
reduce to
\begin{equation}
  \frac{\eta (\xi)}{\eta_0}=
    \begin{cases}
      \dfrac{\sinh[\sqrt{\hat{\sigma}}\xi]}{\sinh[\sqrt{\hat{\sigma}}\xi_0]}
      &  (\xi \leq \xi_0) \\[15truept] 
      \exp[\sqrt{\hat{\sigma}}(\xi_0-\xi)] & (\xi \geq \xi_0),
    \end{cases}\qquad\hbox{(stripes)}\label{etaofxi}
\end{equation}
and
\begin{equation}
  \frac{\eta (\rho)}{\eta_0}=
    \begin{cases}
      \displaystyle \frac{I_1[\sqrt{\hat{\sigma}}\rho]}
      {I_1[\sqrt{\hat{\sigma}}\rho_0]}
      &  (\rho \leq \rho_0) \\
      \noalign{\vskip8truept}
      \displaystyle \frac{K_1[\sqrt{\hat{\sigma}}\rho]}
      {K_1[\sqrt{\hat{\sigma}}\rho_0]}
      &  (\rho \geq \rho_0),
    \end{cases}\qquad\hbox{(caplets)}\label{etaofrho}
\end{equation}
where
\begin{equation}
  \eta_0=
  \begin{cases}
  \left[\sqrt{\hat\sigma}\left(1+\coth\sqrt{\hat{\sigma}}\xi_0\right)
  \right]^{-1}& \textrm{(stripes)} \\[10truept]
  \rho_0 I_1[\sqrt{\hat{\sigma}}\rho_0]
  K_1[\sqrt{\hat{\sigma}}\rho_0]& \textrm{(caplets)}.
  \end{cases}
\end{equation}

\begin{figure}[t]
\includegraphics[scale=0.4]{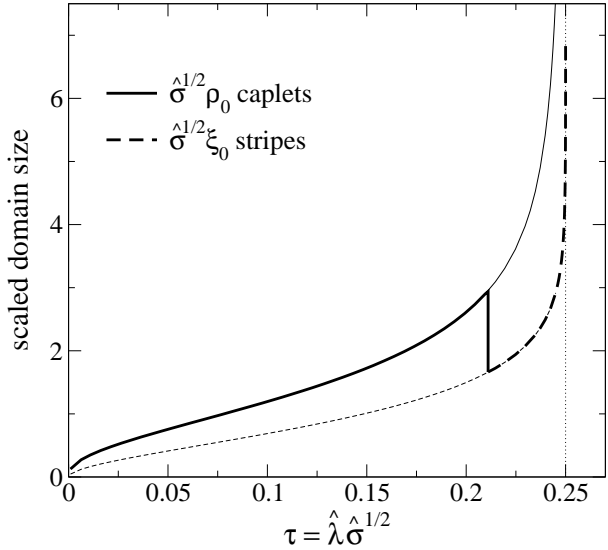}
\caption{
  Reduced domain size as a function of $\tau$ in the dilute limit
  ($\phi\rightarrow0, \hat\mu\rightarrow0$). The solid lines denote the
  stable phase. Caplets are stable for $\tau\lesssim0.2115$, and
  stripes are stable for $0.2115\lesssim\tau<0.25$. For $\tau>0.25$
  the stable state is a macroscopically phase separated flat phase.}
\label{dilutestripesize}
\end{figure}

Substituting these profiles into Eq.\ (\ref{Eq1}), the free energy per
impurity of a single domain is (recall that we set $\hat\mu=0$ in the
dilute limit)
\begin{equation}
  \frac{G_0}{{A^{\perp}_{\rm B}}}=
  \begin{cases}
    \displaystyle\frac{1}{\xi_0}\left[\hat{\lambda} -
      \tfrac{1}{2}\eta_0\right]&\textrm{(stripes)}\\[12truept]
    \displaystyle\frac{1}{\rho_0}\left[2\hat{\lambda} -
      \eta_0\right] & \textrm{(caplets)}, 
  \end{cases}
\end{equation}
where the impurity area ${A^{\perp}_{\rm B}}$ is $2Lx_0$ (stripes) or $\pi r_0^2$ (caplets).  The free energies can be rewritten as
\begin{equation}
  \frac{G_0}{{A^{\perp}_{\rm B}}}=
  \begin{cases}
    \displaystyle
    \frac{\left[\tau- \frac{1}{2}\left(1+\coth\sqrt{\hat\sigma}{\xi}_0\right)^{-2}\right]}
{\sqrt{\hat\sigma}{\xi}_0}
&\textrm{(stripes)}\\ \\%[12truept]
    \displaystyle\frac{\left[2\tau -
      \sqrt{\hat\sigma}\rho_0 I_1\left(\sqrt{\hat\sigma}\rho_0\right) \,
      K_1\left(\sqrt{\hat\sigma}\rho_0\right) \right]}
{\sqrt{\hat\sigma}\rho_0} &  \textrm{(caplets)},
  \end{cases}
\end{equation}
so that the phase diagram in the dilute limit only depends on the
control parameter $\tau\equiv\hat{\sigma}^{1/2}\hat{\lambda}$
\footnote{An analogous stability limit was derived for the model in
  Ref.~ \protect{\cite{Kawa+93}} for stripes in the strong segregation
  limit.}.  Minimizing the free energies over the periods yield the
equilibrium domain sizes, shown in Fig.~\ref{dilutestripesize}. The
caplet phase is stable with respect to stripes for 
$\tau\lesssim0.2115$, while stripes are unstable with respect to a flat
phase for $\tau>0.25$.

%%%%%%%%%%%%%%%%%%%%%%%%%%%%%%%%%%%%%%%%%%%%%%%%%%%%%%%%%%%%%%%%%%%%%%%
%%%%%%%%%%%%%%%%%%%%%%%%%%%%%%%%%%%%%%%%%%%%%%%%%%%%%%%%%%%%%%%%%%%%%%%

The stripe domain size diverges as the limit $\tau=0.25$ is
approached, as
\begin{equation}
  \label{eq:20}
  \lim_{\tau\rightarrow0.25}\xi\hat{\sigma}^{1/2} =
  \frac12\ln\left(\frac{2}{4\tau-1}\right), 
\end{equation}
while the caplet radius grows from zero for small $\tau$ as
\begin{equation}
  \label{eq:23}
  \lim_{\tau\rightarrow 0}\rho\hat{\sigma}^{1/2} =
  \left[\frac{4\tau}{3}\ln\left(\frac{1}{4\tau} \right)\right]^{1/3}.
\end{equation}
Thus, the domain size diverges as either the line tension or the
surface tension becomes large.  The latter case simply corresponds to
the fact that for large surface tensions there will be complete phase
separation in a nearly flat two-dimensional system.  The former case
is somewhat more subtle, although, as we shall see, this limit
corresponds to budding, since the resulting slope $\eta_0$ above
becomes large.  
%%%%%%%%%%%%%%%%%%%%%%%%%%%%%%%%%%%%%%%%%%%%%%%%%%%%%%%%%%%%%%%%%%%%%%%
\subsection{Finite Concentration of the ${\rm B}$ Phase}

%%%%%%%%%%%%%%%%%%%%%%%%%%%%%%%%%%%%%%%%%%%%%%%%%%%%%%%%%%%%%%%%%%%%%%%
For finite impurity concentration of the two phases, the the effective
surface tensions of the two phases differ by the chemical potential
(Lagrange multiplier) $\mu$, which cannot strictly speaking be
ignored. In this case the phase diagram does not depend solely on the
dimensionless quantity $\tau=\hat{\lambda}\hat\sigma^{1/2}$. This was
also noted by  G\'o\'zd\'z and Gompper, in the case of systems with two
different spontaneous curvatures \cite{gozdz01}. However, we will see
that the reduction to a phase diagram that depends only on $\tau$ is
an excellent approximation in many cases.

For finite area fraction $\phi$ of the ${\rm B}$ phase, we assume a
regular array of monodisperse (stable) domains, each within cell
dimensions determined by the concentration.  We approximate the
hexagonal Wigner-Seitz cell (see Fig.~\ref{domains}) by a set of
circular domains of equal area \cite{ziman72}.  The solutions are
given by Eqs.~(\ref{eq:sol1d}) (stripes) and~(\ref{eq:sol2d})
(caplets), with corresponding free energies $\hat{g}$ given by
Eqs.~(\ref{eq:fstripe}) and~(\ref{eq:fcap}).  To calculate the phase
diagram we minimize the free energies over both domain dimensions
($x_0$ and $X_0$ for stripes, $r_0$ and $R_0$ for caplets) for a given
set of control parameters (line tension $\hat{\lambda}$, frame tension
$\hat{\sigma}$, chemical potential $\hat{\mu}$), and determine the phase
boundaries by that chemical potential $\hat{\mu}$ for which the free
energies of the caplet and stripe phases are the same.  

\begin{figure}[t]
\includegraphics[width=3.0truein]{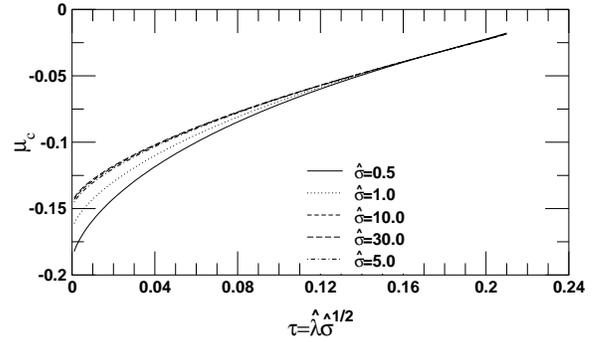}
\caption{Chemical potential as a function of generalized tension for
  different frame tensions $\hat{\sigma}$. Note the extrapolation to small
  $\hat{\mu}$ with increasing $\tau$.}
\label{fig:mu}
\end{figure}

Our procedure differs slightly from that of Ref.~\cite{HardMack94} in
that, here, the effective overall ``tension'' is different in the two
phases. In the impurity phase $B$ the quantity $\hat{\sigma}+\hat{\mu}$ acts
like a mechanical tension penalizing area changes, while (because we
have chosen a reference chemical potential $\hat{\mu}_A=0$) only
$\hat{\sigma}$ penalizes area changes in the $A$ phase. Hence the inverse penetration
lengths ${\zeta}_A$ and ${\zeta}_B$ differ.  In the strong tension
limit $\hat{\sigma}\gg\hat{\mu}$, which is exact for vanishing $\phi$ where
$\hat{\mu}$ becomes quite small (see Fig.~\ref{fig:mu}), this approach
recovers the results of Ref.~\cite{HardMack94}, while for smaller
$\hat{\sigma}$ the height profile of the impurity phase contains a
contribution from the chemical potential as well as the frame tension.
In the limit $\hat{\sigma}\gg\hat{\mu}$ where the penetration depths are the
same, the phase diagram depends on only the generalized reduced
tension ${\tau}$.

%%%%%%%%%%%%%%%%%%%%%%%%%%%%%%%%%%%%%%%%%%%%%%%%%%%%%%%%%%%%%%%%%%%%%%%
%%%%%%%%%%%%%%%%%%%%%%%%%%%%%%%%%%%%%%%%%%%%%%%%%%%%%%%%%%%%%%%%%%%%%%%
%%%%%%%%%%%%%%%%%%%%%%%%%%%%%%%%%%%%%%%%%%%%%%%%%%%%%%%%%%%%%%%%%%%%%%
\begin{figure}[ht]
  \includegraphics[scale=0.43]{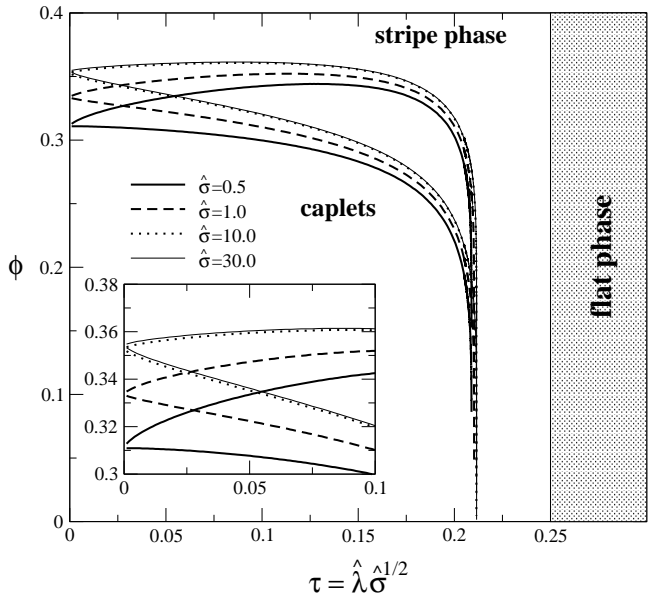}
\caption{Phase diagram as a function of
  $B$-phase area fraction $\phi$ and generalized reduced tension
  ${\tau}$, for different constant frame tensions $\hat{\sigma}$
  ($\hat{\lambda}$ was varied to scan $\tau$).  The region between
  upper and lower pairs of lines are two-phase regions of coexisting
  stripes and caplets. The inset shows convergence for large
  $\hat{\sigma}$ near the small $\tau$ phase boundary. An equilibrium
  flat phase is predicted for ${\tau}>0.25$.  Scans for different
  $\hat{\sigma}$ converge to the results of Ref.~\cite{HardMack94},
  which corresponds to the limit $\hat{\sigma}\rightarrow\infty$.}
\label{fig:diag}
\end{figure}
%%%%%%%%%%%%%%%%%%%%%%%%%%%%%%%%%%%%%%%%%%%%%%%%%%%%%%%%%%%%%%%%%%%%%%%

With increasing $\phi$, the domains grow monotonically with
$r_0(\phi)\simeq2\xi_0(\phi)$, which corresponds to nearly equal mean
curvature $c\sim c_0$.  Note that a finite preferred domain size
implies the existence of an equilibrium mesophase.  With increasing
$\phi$ there is a first order transition from circular caplets to
stripes for small $\tau$ (Fig.~\ref{fig:diag}).  For small $\tau$ the
transition depends very weakly on $\tau$ and coexistence occurs over a
range of concentrations near $\phi =0.35$.  This range moves to
smaller concentration with increasing $\tau$, while for $0.2115 <
{\tau} < 0.25 $, however, the stripe phase becomes stable at all
concentrations.

\begin{figure}[ht]
\includegraphics[width=3.0truein]{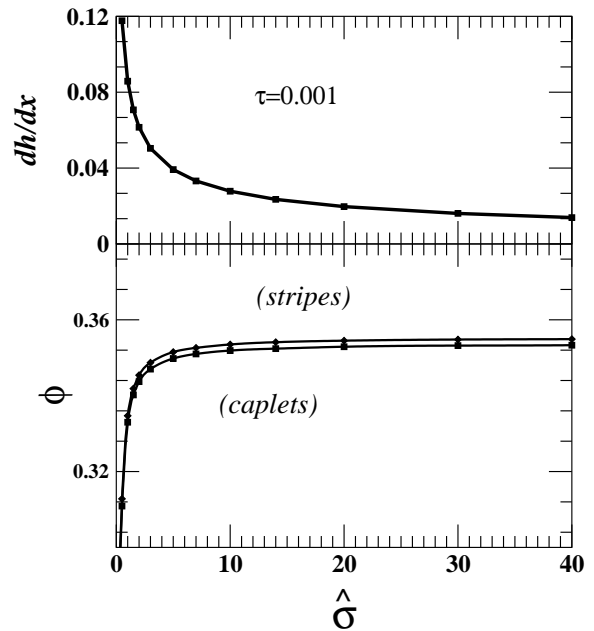}
\caption{Phase boundaries $\phi_{\rm caplet}$ and $\phi_{\rm stripe}$,
  and stripe and caplet slopes $dh/dx$ for fixed $\tau=0.001$ and varying
  $\hat{\sigma}$.}
\label{fig:slopes}
\end{figure}

Figs.~\ref{fig:mu}, \ref{fig:diag}, and \ref{fig:slopes} show that for
large $\hat{\sigma}$ the phase diagrams converge to a scaling form
that depends only on $\tau$, recovering the phase diagram calculated
in Ref.~\cite{HardMack94}. For larger $\hat{\sigma}$ the concentration
range of the stable caplet phase widens at small $\tau$, moving to
larger $\phi$.  Recall, Eq.~(\ref{eq:24}), that the characteristic
inverse decay length in the impurity phase is
$\zeta_B=\sqrt{\hat{\sigma} + \hat{\mu}}$, where we find $\hat{\mu}<0$
along the phase boundary.  Hence, for a smaller tension $\hat{\sigma}$
the negative potential decreases the energetic cost of buckling
against tension, which destabilizes the stripe phase.  The deviation
of the phase boundary near $\tau\simeq0.2115$ is due to departures
from the Monge limit.

%%%%%%%%%%%%%%%%%%%%%%%%%%%%%%%%%%%%%%%%%%%%%%%%%%%%%%%%%%%%%%%%%%%%%%%
%%%%%%%%%%%%%%%%%%%%%%%%%%%%%%%%%%%%%%%%%%%%%%%%%%%%%%%%%%%%%%%%%%%%%%%

\begin{figure}[ht]
\includegraphics[width=3.0truein]{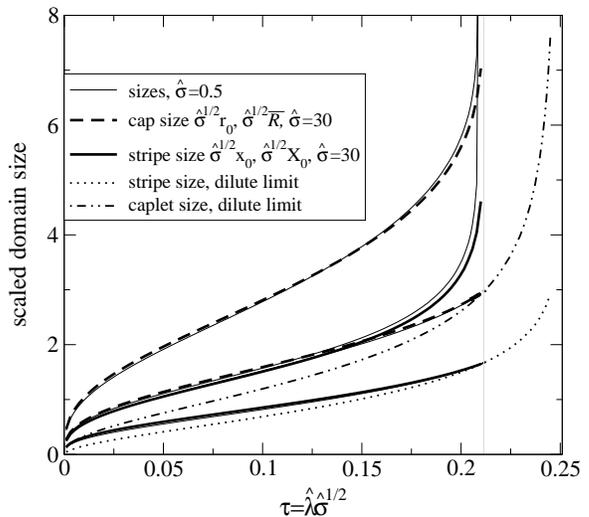}
\caption{Scaled domain sizes for caplet and stripe phases along the
  phase coexistence line, for both the concentrated and dilute limit
  calculations. For each geometry the two dimensions (stripe
  half-period $\xi_0$ and minority domain half width $\Xi_0$, and caplet
  inner radius $\rho_0$ and Wigner-Seitz radius $\bar{R}$, all scaled by
  $\hat{\sigma}$) are shown. The variation in domain sizes with
  tension $\hat{\sigma}$ can be seen to be very small.}
\label{fig:sizes}
\end{figure}

The length scales in the concentrated limit are larger than in the
dilute limit, for the same $\tau$, and converge to the dilute limit
result when the caplet phase loses stability at $\tau\simeq0.2115$, as
expected (Figure \ref{fig:sizes}).

%%%%%%%%%%%%%%%%%%%%%%%%%%%%%%%%%%%%%%%%%%%%%%%%%%%%%%%%%%%%%%%%%%%%%%%

%%%%%%%%%%%%%%%%%%%%%%%%%%%%%%%%%%%%%%%%%%%%%%%%%%%%%%%%%%%%%%%%%%%%%%%
\section{Numerical solutions in the budding regime}\label{sec:num}
%%%%%%%%%%%%%%%%%%%%%%%%%%%%%%%%%%%%%%%%%%%%%%%%%%%%%%%%%%%%%%%%%%%%%%%
The calculations presented above and in Ref.~\cite{HardMack94} are
based on a small-slope approximation, and are only valid for small
tensions $\sigma$. In order to both test the validity of this
approximation as well as to study the possible transition from the
caplet state above to a budded states, we have performed direct
numerical minimization of the model free energy in Eq.~(\ref{Eq1}),
beyond the small-slope approximation. We present here the equilibrium
shapes and phase diagram for the stripe phase, the (azimuthally
symmetric) caplet phase, and for buds.

We focus on the dilute limit, i.e., for small area fraction $\phi$. It
is expected that phase separation may, under appropriate conditions,
lead to the formation of buds (nearly spherical domains of one phase
connected with a narrow neck to another phase that is flat). It has
previously been shown theoretically that the line tension between two
phases alone may result in budding \cite{Lipo92b,JuliSeif94}. The role
of phase separation in budding and fission has also been studied in
recent experiments \cite{Miao+94}. As described above, our small-slope
approximation used in the previous sections is expected to fail for
large line tension $\lambda$.

In order to address both the validity of the preceding analysis, as
well as to treat possible budding transitions, we describe the
membrane shape by a local tangent angle $\theta$ relative to a
horizontal, flat conformation. For a stripe domain, the shape can be
completely characterized by the tangent angle as a function of only
one contour length coordinate $s$, which is defined along a line
perpendicular to the stripe. For a caplet or bud domain, a single
tangent angle as a function of a single coordinate $s$ can also be
used to completely describe the membrane shape, even in the presence
of overhangs.  This can be done provided that the domain is symmetric
about an axis perpendicular to a horizontal plane. In this case, the
contour length coordinate is defined along a radial direction. A
similar approach was used in \cite{gozdz01}.

%%%%%%%%%%%%%%%%%%%%%%%%%%%%%%%%%%%%%%%%%%%%%%%%%%%%%%%%%%%%%%%%%%%%%%%
\begin{widetext}
  \begin{center}
\begin{figure}[h]
    \includegraphics[width=6.0truein,bb=65 30 740 520]{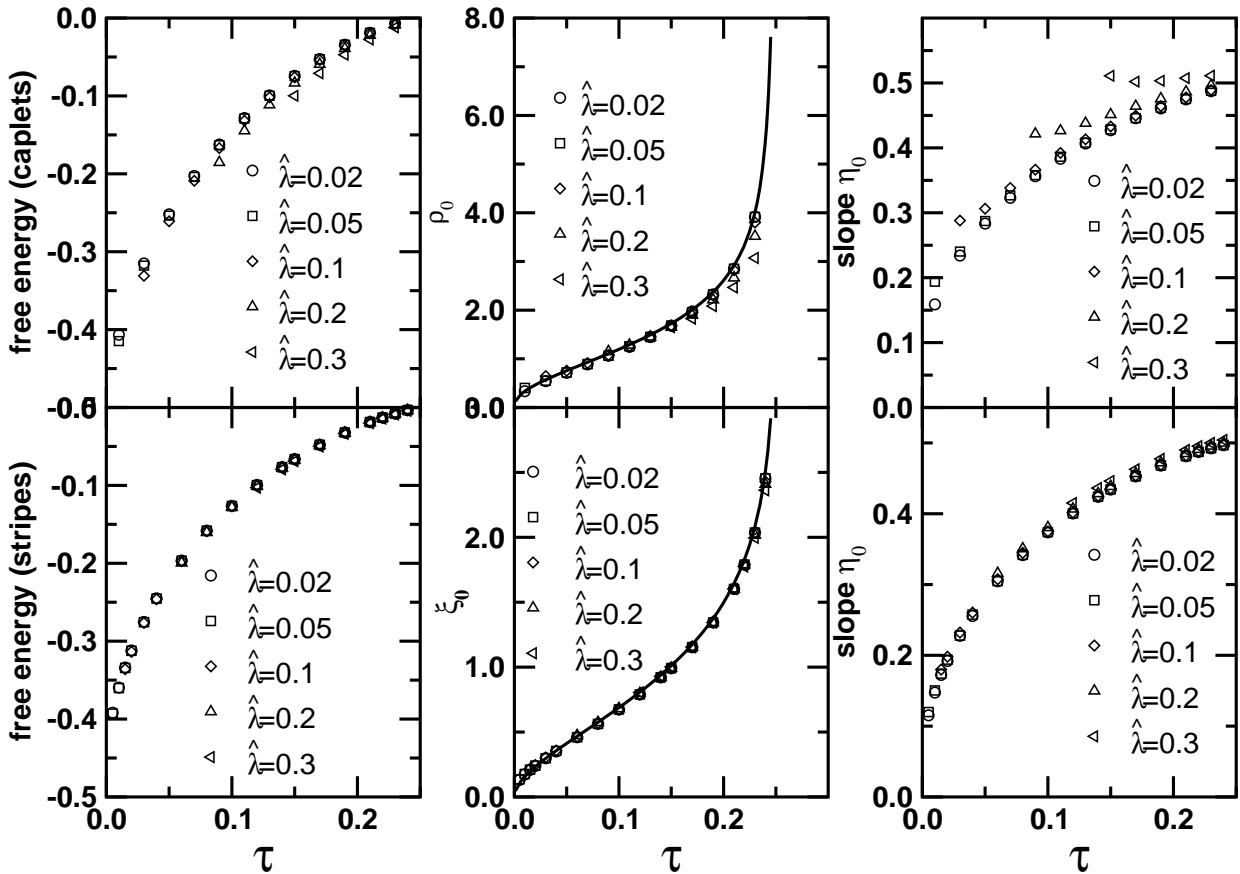}
\caption{Data collapse of the free energy per unit area (in units of
  $\kappa c_0^2$), domain size (units of $c_0$) and boundary slope as
  a function of $\tau$ for both stripes and caplets, for various values
  of the line tension $\hat\lambda$.}
\label{scaling}
\end{figure}
  \end{center}
\end{widetext}
%%%%%%%%%%%%%%%%%%%%%%%%%%%%%%%%%%%%%%%%%%%%%%%%%%%%%%%%%%%%%%%%%%%%%%%
If the center of the domain is taken as the origin, then the membrane
shape in three dimensions is given by
\begin{equation}
\left(x(s),y(s),z(s)\right)
=\left(r(s)\cos(\phi),r(s)\sin(\phi),z(s)\right),
\end{equation}
where $r(s)=\int\cos(\theta)\;ds$ is a radial coordinate,
$z(s)=\int\sin(\theta)\;ds$, and $\phi$ is the azimuthal angle.
In this coordinate system, the mean curvature is given by
\begin{equation}
H=\frac{d\theta}{ds}+\frac{\sin\theta}{r}
\end{equation}
for caplet and bud domains. For stripes, only the first term above is
necessary.

The resulting free energy is evaluated numerically for a domain shape
defined by a discrete set of 50 to 200 angles
$\{\theta_i\}_{i=0,\ldots,N}$, which are defined at equally-spaced
points $s_i$ along the contour from the origin to the Wigner-Seitz
boundary described previously. These values, together with an overall
scale factor $\Delta s=s_{i+1}-s_i$ determine the membrane shape.

A discrete approximation to the free energy $f$ per unit area of
domain is minimized to determine equilibrium shape and free energy.
Here, we quote results for an area fraction of 10\%. As shown above
within the small-slope approximation, the phase boundary between
stripes and caplets is very narrow and relatively insensitive to area
fraction in this range. In Fig.\ \ref{scaling}, we show the normalized
stripe domain size ($\xi_0$), boundary slope
($\eta_0$), and free energy $f$ vs $\tau$ for several
different combinations of $\hat\sigma$ and $\hat\lambda$.  The
apparent collapse of the data demonstrates one of the conclusions of
the previous analysis, namely that the results can be well-represented
in terms of a single combination of parameters,
$\tau=\hat\lambda\sqrt{\hat\sigma}$, of surface tension and line
tension. Similar results for caplet domains are shown in Fig.\ 
\ref{scaling}.  Here, too, the results show the dependence on the
single parameter $\tau$.

\begin{figure}
\includegraphics[width=2.8truein,angle=0]{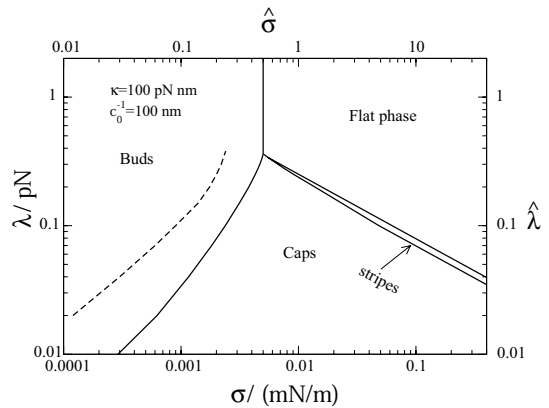}
\caption{The calculated phase diagram, indicating bud, caplet, and stripe
  phases, as well as the flat phase corresponding to macroscopic phase
  separation. This has been obtained by the numerical procedure
  outlined in Sec.\ \ref{sec:num}. The solid lines indicate
  thermodynamic phase boundaries, while the dashed line indicates the
  limit of metastability of caplets. As discussed in the text, we
  expect that buds are metastable throughout the phase diagram, within
  the approximations used here.  The thin region shown is that of the
  stripe phase. This vanishes at a triple point near $(\hat{\sigma},
  \hat\lambda)=(1,0.2)$. The bud, caplet, and flat phases meet at the
  other triple point along $\hat{\sigma}=0.5$.  Upper and right hand
  axes are dimensionless, while lower and left hand axes are in
  physical units for typical values of the bending energy and
  spontaneous curvature, $\kappa=100\,\textrm{pN nm},
  c_0^{-1}=100\,\textrm{nm}$.}
\label{regimes}
\end{figure}

These numerical results can also be used to determine the phase
boundary between the caplet and stripe phases. For simplicity, we do this
only in the dilute limit, in which the phase boundary is rather
insensitive to concentration. Specifically, we find the crossing of
the free energies of the caplet and stripe phases ($f_c$ and $f_s$) for
10\% area fraction.  The results are shown in Fig.~\ref{regimes}. We
note the excellent agreement of this phase boundary with that shown in
Fig.~\ref{fig:diag}, at least for $\hat\lambda\alt 0.2$. For larger
$\hat\lambda$ the former approximations fail. One consequence of this
is that the stripe phase vanishes with increasing $\hat\lambda$:
specifically, there is a narrow range of values of the line tension
for which the only stable domains are caplets. This can be seen in Fig.\ 
\ref{regimes}. The simple physical reason for the enhanced stability
of the caplet phase (compared with the small-slope results above) in this
limit is that the line tension cost is lowered in the caplet geometry for
large slopes, given a fixed domain area: for caplets, the circumference
to area ratio is reduced with increasing slope.

Also shown in Fig.~\ref{regimes} is the transition between caplet and
bud states, as well as a spinodal line beyond which no stable caplet
structures are possible. No spinodal representing the limit of
metastability of buds is found. This is an artifact of the use of a
bending free energy using only leading order terms. Such a spinodal
depends on the details, and especially the radius $a$ of the neck in
the budded state. Without further assumptions, this length shrinks to
of order $\Delta s$. The surface tension required to pull-out such a
bud is of order the two dimensional Laplace ``pressure''
${\sigma}\sim{\lambda/a}$, which diverges for small necks. Thus, this
spinodal cannot be reliably determined, if at all, by our numerical
scheme. Within the approximation that the neck of the budded state
shrinks to a size much smaller than the domain size, we find that buds
are, in fact, metastable throughout the phase diagram. This is
because, for small necks, there is a linear increase in the free
energy due to the line tension when the neck expands, while both
bending and surface tension contributions are quadratic.

The remaining phase boundary shown in
Fig.~\ref{regimes} is that of the transition from
the flat phase (macro phase separation) to the budded state. This
phase boundary is simple to estimate. The free energy per unit
area of the bud is simply $-\tfrac12\kappa c_0^2$, while the bud
accounts for an excess area of $4\pi\left(2/c_0\right)^2$,
assuming an optimal mean curvature of $c_0$. As noted above, the
radius of the neck tends to zero. This results in a phase
boundary given by $\hat\sigma=1/2$ in reduced units. This simple
phase boundary estimate is well borne out by the numerical
minimization, as shown in the figure.
%%%%%%%%%%%%%%%%%%%%%%%%%%%%%%%%%%%%%%%%%%%%%%%%%%%%%%%%%%%%%%%%%%%%%%%
%%%%%%%%%%%%%%%%%%%%%%%%%%%%%%%%%%%%%%%%%%%%%%%%%%%%%%%%%%%%%%%%%%%%%%%

%%%%%%%%%%%%%%%%%%%%%%%%%%%%%%%%%%%%%%%%%%%%%%%%%%%%%%%%%%%%%%%%%%%%%%%
\section{Discussion}\label{sec:yadayada}
%%%%%%%%%%%%%%%%%%%%%%%%%%%%%%%%%%%%%%%%%%%%%%%%%%%%%%%%%%%%%%%%%%%%%%%
The transition from caplets to stripes with increasing $\phi$ and
$\tau$ can be understood in simple physical terms.  On the one hand,
with increasing $\phi$ the unit cells in the stripe and caplet
morphologies are of size $\sim 1/\phi$ and $\sim 1/\sqrt{\phi}$,
respectively.  Thus, with increasing $\phi$, the constraints on the
membrane profiles due to the unit cell boundaries become more costly
in the caplet phase than the stripe phase.  On the other hand, with
increasing tension $\tau$, the regions of largest slope (i.e., largest
difference between projected area and membrane area), which occur near
the domain boundaries, become energetically more costly because of the
work required against tension.  For caplets, this represents a larger
fraction of the domain area than for stripes.  Thus, stripes are
preferred for sufficiently large $\tau$.

From the numerical approach of the previous section, we see, perhaps
surprisingly, that the Monge approximation works rather well to
describe the various phases and transitions. Specifically, we see from
Fig.\ \ref{scaling} that for $\hat\lambda\alt 0.2$ there is little
disagreement between the Monge approximation and the numerical
solutions that can account for finite slopes. This is particularly
true of the caplet phase.

In general, we expect the Monge approximation to be best for large
$\hat\sigma$, which tends to flatten the membrane. Since the phase
boundaries between caplets, stripes, and the flat phase tend to occur for
a fixed $\tau=\hat\sigma^{1/2}\hat\lambda$ (specifically
$\tau\simeq0.21$ and $\tau=0.25$ for the caplet-stripe and stripe-flat
boundaries, respectively, in the dilute limit), we expect the Monge
approximation to be valid near both transitions, except for large
$\lambda$, which is what we find. Here, of course, $\sigma$ tends to be
smaller.

As noted above, however, the discrepancy is smaller for the caplet
phase.  The deviation that we find for stripes suggests an instability
toward larger domains and, ultimately, the flat phase. This tends to
be compensated in the case of caplets by the line tension $\lambda$,
which prevents domains from growing. This enhanced stability of
caplets is also consistent with the observation from the phase diagram
in Fig.~\ref{regimes} that the caplet-stripe phase boundary shifts in
favor of caplets as $\lambda$ increases. We find, in fact, a triple
point, where caplet, stripe and flat phases meet, although this is
hard to see in the figure.  For larger $\lambda$, no stripe phase
is observed.

Away from these transitions, however, we begin to see significant
deviation from the Monge results for the largest $\lambda$ as sigma
decreases. Thus, the Monge approximation tends to fail to the left in
Fig.\ \ref{regimes}. In particular, the Monge approximation is
insufficient to characterize the transition between buds and caplets.

This work shares some common features with several previous
theoretical studies.  Undulating stripe and hexagonal phases were
predicted in the limit of weakly segregated two-component films
\cite{leibler87} based on a phenomenological coupling of local
membrane composition and shape.  These phases are analogous to the
stripe and caplet phases discussed in this letter for well-segregated
mixtures.  The approach of Ref.\ \cite{leibler87} has recently been
utilized to study one-dimensional shape profiles of two-component
membranes and vesicles in the strong segregation regime as well
\cite{AKK92,Kawa+93}.  The stripe phases of planar membranes in the
limit of small domains or large bending energies reported in
Ref.\cite{Kawa+93} are qualitatively similar to our stripe mesophases.
The stable stripe and caplet mesophases reported in this letter should
occur in phase-separated two-component films under tension, provided
that one phase has finite spontaneous curvature and that the line and
surface tensions are sufficiently small.  In particular, one must have
$\tau<1/4$ and $\hat\lambda <\hat\sigma$ to observe these mesophases.
The calculation presented above assumes an asymptotically flat matrix
phase, as in the case of a globally flat monolayer or bilayer
membrane.  However, our results should be qualitatively valid for
two-component bilayer vesicles, provided that the vesicle diameter $D$
is sufficiently large compared with the characteristic domain
dimension, \emph{i.e.}, if $c_0D\gg 1$.

G\'o\'zd\'z and Gompper performed a similar calculation to ours in a
different ``slice'' of parameter space. They considered phases with
\textit{different} signs of spontaneous curvature, and predicted a
variety of different budding, stripe, and circular domains. It is
difficult to compare the results directly, since they chose only one
parameter set, $\hat{\lambda}=0.5, \hat{\sigma}=0.4$. For our model,
with just one spontaneous curvature, this leads to a budded state.
They show that this set of parameters can lead to a variety of phases
that can be stabilized by the different spontaneous curvatures.

We estimate the region of domain stability for the following values of
the parameters in our model:
\begin{align}
  \kappa&= 25 k_BT\simeq 100 \textrm{pN nm},& c_0^{-1}&=100 {\rm
    nm}.\label{eq:numbers} 
\end{align}
The results are shown schematically in Fig.~\ref{regimes}.  Thus,
caplet domains of order $100\,\textrm{nm}$ in size (as in Ref.\ 
\cite{Baek+95}) are stable for line tensions
$\lambda\alt0.3\,\textrm{pN}$ and surface tensions
$\sigma\sim10^{-2}\,\textrm{mN/m}$.  For comparison, membrane rupture
occurs for surface tensions of order $10 \textrm{mN/m}$
\cite{biomembranes}.

Finally, we wish to note a couple of important general observation
relevant to phase separating membrane systems.  First, given the
coupling between membrane composition and shape, it is possible to
have both composition-induced changes in shape or curvature
\cite{leibler86,JuliLipo93,MackSafr93}, as well as curvature-induced
changes in composition \cite{Seif93}. However, unless the curvature is
of a molecular scale (\emph{i.e.}, if curvature radii are comparable
to the size $a$ of a single lipid), it is generally expected that
composition will drive shape, or, in other words, that chemistry
determines geometry. This is because, on a per molecule basis, the
curvature energy $\kappa a^2c_0^2$ is very small compared with $kT$.
Thus, unless one happens to be very close to a critical point
\cite{Seif93}, the shift in chemical potential due to the local
curvature of the membrane is expected to be insufficient to cause a
significant composition change.  Second, the Gaussian curvature and
associated modulus $\bar\kappa$, which we have so far neglected, is
expected to affect the stability of the caplet and budded phases
described above. In fact, this can have an order one effect on both
phases, since the corresponding shift on the curvature energy is of
order $\Delta\bar\kappa c_0^2$ per unit area, where $\Delta\bar\kappa$
represents the difference in moduli between the two phases. This is
comparable to the (mean) curvature energy contribution in both phases.
There will be no effect, however on the stripe phase free energy.
Thus, the main qualitative effect of taking into account the Gaussian
curvature effects will be on the stability of the bud and caplet
phases relative to the stripe phase. For instance, the expected
increase or decrease in the energy of bud and caplet phases may result
in a more pronounced or even absent stripe. Finally, we have imposed
the strong segregation limit, in which the compositions of the two
phases are unaffected by the local bending.  This is likely to break
down near critical points, at which point the compositions will vary
in concert with the degree of bending
\cite{leibler87,safranpinc90,safranpinc91,AKK92,Kawa+93,Tani+94}. In
this case the local line tension can be expected to be lower and
depend on the degree of local membrane deformation. Moreover, elastic
constants and the spontaneous curvature will be inhomogeneous in the
more highly deformed regions. Possible effects of this could be to
stabilize the caplet phase in favor of the stripe phase (since line
tension may be reduced), as well as shifting other phase boundaries.
However, as we have emphasized above, the strong segregation limit is
expected to apply more generally, because of the wide separation of
scales between the molecular and curvature lengths.

%%%%%%%%%%%%%%%%%%%%%%%%%%%%%%%%%%%%%%%%%%%%%%%%%%%%%%%%%%%%%%%%%%%%%%%
\section*{Acknowledgements}
%%%%%%%%%%%%%%%%%%%%%%%%%%%%%%%%%%%%%%%%%%%%%%%%%%%%%%%%%%%%%%%%%%%%%%%
We thank A.~Ajdari, F.~J\"ulicher, D.~Morse, J.~Prost, R.~Ross, and
E.~Sackmann for useful discussions and comments.  FCM acknowledges
support from the NSF (Grant No. DMR-9257544), the Donors of the
Petroleum Research Fund, and the Exxon Education Foundation. PDO and
FCM acknowledge NATO CRG 960678.  All three authors thank the (K)ITP
UCSB and the Isaac Newton Institute for Mathematical Sciences, where
portions of this work were also performed. The authors wish to thank
the ICTPC for a stimulating environment.
%%%%%%%%%%%%%%%%%%%%%%%%%%%%%%%%%%%%%%%%%%%%%%%%%%%%%%%%%%%%%%%%%%%%%%%
\appendix
%%%%%%%%%%%%%%%%%%%%%%%%%%%%%%%%%%%%%%%%%%%%%%%%%%%%%%%%%%%%%%%%%%%%%%%
\section{Derivation of Free Energy in Monge Gauge}\label{app:a}
%%%%%%%%%%%%%%%%%%%%%%%%%%%%%%%%%%%%%%%%%%%%%%%%%%%%%%%%%%%%%%%%%%%%%%%%%%%%%
In this appendix we derive the free energy and boundary conditions for
the Monge gauge calculation.
%%%%%%%%%%%%%%%%%%%%%%%%%%%%%%%%%%%%%%%%%%%%%%%%%%%%%%%%%%%%%%%%%%%%%%%%%%%%%
\subsection{Free Energy Variation}
%%%%%%%%%%%%%%%%%%%%%%%%%%%%%%%%%%%%%%%%%%%%%%%%%%%%%%%%%%%%%%%%%%%%%%%%%%%%%
Consider an impurity phase arranged into domains. The free energy is
given by Eq.~(\ref{Eq1}). The bending free energy with respect to a
flat reference state is given by
\begin{equation}
F_b = \frac12\kappa \int_{S_{\rm B}}\!dS
 \left[c\left(r\right)^2 - 2c\left(r\right) c_0+ c_0^2\right]
+ \frac12\kappa \int_{S_{\rm A}}\!\!dS\, c(r)^2,
\end{equation}
where $c(r)$ is the mean curvature, $c_0$ is the spontaneous curvature
of the impurity phase, $\kappa$ is the bending modulus, and $dS$ is
the surface area.  The structure has a frame tension ${\sigma}$ against
which the membrane does work, with energy
\begin{equation}
F_s =   \sigma \left\{A_{fr} - \int_{S_{\rm A} + S_{\rm B}}
  dS_{\perp}\right\},
\end{equation}
where $dS_{\perp}$ is the projected membrane area and $A_{fr}$ is the
frame area. The line tension energy is proportional to the interface
between phases ${\rm A}$ and ${\rm B}$ and is given by
Eq.~(\ref{Eq2}).

The membrane composition  $\phi$ is given by
\begin{equation}
\phi = \frac{S_{\rm B}}{{S_{\rm B}} +S_{\rm A}}.
\end{equation}
The two areas $S_{\rm A}$ and $S_{\rm B}$ are determined by the frame
tension and a Lagrange multiplier $\mu$ to control the impurity phase
$B$.  Since we assume a constant area per lipid, $\mu$ is proportional
to the chemical potential for $B$, or more strictly the exchange
chemical potential difference between $A$ and $B$.  We must minimize
\begin{equation}
G_0(\mu,\sigma,\lambda) = F_b + F_s + F_l + \mu
\int_{S_{\rm B}}dS 
\label{eq:ftot}
\end{equation}
over variations in the surface shape, $h(r) \rightarrow h(r) + \delta
h(r)$, where the line tension energy $F_l$ is given by
Eq.~(\ref{Eq2}). We will later minimize the resulting free energy over
the domain dimensions.

In the Monge gauge in the $2d$ geometry of the caplet case,
\begin{subequations}
\begin{align}
c(r) &= \nabla^2 h =\frac12\frac{\partial}{\partial
  r}\left(r\frac{\partial h}{\partial r}\right)\\
dS &= \left[1 + \frac12 \left(\nabla h\right)^2\right]^{1/2}d^2\!r \\
dS_{\perp} &= d^2\!r,
\end{align}
\end{subequations}
and the variation of $G$ over $h$ leads, to lowest order in $h$,
\begin{widetext}
\begin{align}
  \frac{\delta G}{2\pi} &=
  \delta h(r_0) \, f_{hr}(r_0) + \delta h(R) \, f_{hR}(R) +\,\,
  \delta h'(r_0) \, \Gamma_{r}(r_0)
  + \delta h'(R) \Gamma_{R}(R)
  +  \int_0^{r_0}\!\!\!d^2r \,\delta h f_{\it e-l}^{\rm B}
  + \int_0^{R}\!\!\!d^2r\,\delta h f_{\it e-l}^{\rm A},\label{eq:balance}
\end{align}
\end{widetext}
in terms of the following forces $f$ and torques $\Gamma$:
\begin{subequations}
\begin{align}
f_{\it hr}&=2\pi\, r\left[\kappa\left({c}_{\rm A}' -
{c}_{\rm B}'\right) -(\sigma+\tilde{\mu}) h_{\rm B}' + \sigma
h_{\rm A}' \right] \label{eq:bc3}\\
f_{\it hR}&=2\pi\, R\left[\sigma
h_{\rm A}' - \kappa {c}_{\rm A}' \right] \label{eq:bc4} \\
\Gamma_{r}&=2\pi\,\kappa r \left(  {c}_{\rm B} -
{c}_{\rm A} - c_0\right) \\
\Gamma_{R}&=2\pi\,R \kappa{c}_{\rm A}.
\label{eq:bc6}
\end{align}
\label{eq:bcs}
\end{subequations}
Here we have defined $\tilde{\mu}=\mu+\kappa c_0^2$, which is
consistent with a complete expansion to second order in $h$
\cite{gozdz01}.  The Euler-Lagrange equations, which determine
the profile $h(r)$ are
\begin{subequations}
\begin{align}
f_{\it e-l}^{\rm B} &= \kappa ({c}'_{\rm B} + r{c}''_{\rm B}) -
(\sigma+\tilde{\mu}) r {c}_{\rm B} =0 \\
f_{\it e-l}^{\rm A} &= \kappa ({c}'_{\rm A} + r{c}''_{\rm A}) -
\sigma r {c}_{\rm A} =0.
\end{align}
\label{eq:euler2d}
\end{subequations}
Because the free energy and the forces and torques depend only on
derivatives of $h$, we will solve for the slope $\eta=dh/dr$ instead
of the height profile.

Equations~\ref{eq:bcs}-\ref{eq:euler2d} are specific for the
cylindrical geometry of the caplet morphology. The analogous set of
equations for the stripe morphology is:
\begin{subequations}
\begin{eqnarray}
  \frac{f_{\it hr}}{L} &=& \kappa\left({c}_{\rm A}' -
    {c}_{\rm B}'\right) -(\sigma+\tilde{\mu}) h_{\rm B}' +
  \sigma  h_{\rm A}'  \\ 
  \frac{f_{\it hR}}{L} &=&  \sigma
  h_{\alpha}' - \kappa {c}_{\rm A}'  \\
  \frac{\Gamma_{r}}{L} &=&\kappa\,\left({c}_{\rm B}-{c}_{\rm A} - c_0\right) \\
  \frac{\Gamma_{R}}{L} &=& \kappa\,{c}_{\rm A} \\
  f_{\it e-l}^{\rm B}&=&\kappa\, {c}''_{\rm B}-(\sigma+\tilde{\mu}) c_{\rm B} =0 \\
  f_{\it e-l}^{\rm A}&=&\kappa\, {c}''_{\rm A} - \sigma c_{\rm A} =0,
\end{eqnarray}
\label{eq:bceuler1d}
\end{subequations}
where now $h$ is a function of the cartesian coordinate $x$, and the
curvature is given by $c=d^2h/dx^2$.
%%%%%%%%%%%%%%%%%%%%%%%%%%%%%%%%%%%%%%%%%%%%%%%%%%%%%%%%%%%%%%%%%%%%%%%%%%%%%
\subsection{Boundary Conditions}\label{sec:BC}
%%%%%%%%%%%%%%%%%%%%%%%%%%%%%%%%%%%%%%%%%%%%%%%%%%%%%%%%%%%%%%%%%%%%%%%%%%%%%
The boundary conditions are specified by the forces and torques
applied to the membrane. There is no applied torque at the interface,
$r_0$, and we assume no applied torque at the boundary
\footnote{Alternatively, we could specify the slope at the boundary
  $R$.}, so
\begin{subequations}
\begin{eqnarray}
\Gamma_{r}(r_0) &=& 0 \\
\Gamma_{R}(R) &=& 0.
\end{eqnarray}
\label{eq:torques}
\end{subequations}
In this Appendix we will explicitly describe the boundary conditions
for the two dimensional caplet geometry; the same conditions hold for the
one dimensional stripe geometry, with $r\rightarrow x_0$, unless
noted.  The second condition specifies zero curvature on the boundary
while the first, first given by Kozlov and Helfrich, relates the
difference in the mean curvature to the spontaneous curvature of the
inner surface. Next, we assume there are no vertical forces applied to
the membrane,
\begin{subequations}
\begin{eqnarray}
f_{\it hr}(r_0) &=& 0 \label{eq:hr}\\
f_{\it hR}(R_0) &=& 0.\label{eq:hR}
\end{eqnarray}
\label{eq:vertical}
\end{subequations}
However, it is straightforward to show (for both the one and two
dimensional cases) that these vertical boundary conditions are in fact
proportional to first integrals of the Euler-Lagrange equations.
Specifically, in two dimensions,
\begin{subequations}
\begin{eqnarray}
f_{\it h}(r) &=& r\left( \kappa {c}'(r) - \sigma h'(r)\right) \\
&=& \int \!dr\,f_{\it e-l} \left[h(r)\right] \\
&=& \hbox{constant}.
\end{eqnarray}
\end{subequations}
Since $f_{\it h}(r)$ is constant for all $r$, the boundary condition
at the junction, Eq.~(\ref{eq:hr}), is irrelevant and we are left with
Eq.~(\ref{eq:hR}). A similar result applies for the one dimensional
stripe geometry.

To completely specify the problem we need a few more conditions.
Symmetry about the domain center requires
\begin{equation}
h'_{\rm B}(0) = h'''(0) = 0.
\label{eq:h2}
\end{equation}
Another condition on the derivative of the profile $h'(r_0)$ results
by requiring a well-defined energy. If $h'$ has a discontinuity at the
boundary, then the curvature has a delta function as the boundary is
crossed. This leads to a non-physical and singular energy in the
region of the junction $r=r_0$. Hence we are led to our final boundary
condition,
\begin{equation}
h_{\rm B}'(r_0) = h_{\rm A}'(r_0).
\label{eq:h4}
\end{equation}

For completeness, we reproduce the relevant boundary conditions that
remain after recognizing the redundance of Eq.~(\ref{eq:hr}).  For
the single-domain problem, from Eqs.~(\ref{eq:bc4}-\ref{eq:bc6}) we
have:
\begin{subequations}
\begin{eqnarray}
%f_r(r_0)&=& 0 \label {eq:boundary1}\\
\left[{c}_{\rm B} - {c}_{\rm A}\right]_{r_0} &=&
c_0\label{eq:boundary2}\\ 
\tau h_{\rm A}'(\infty) - \kappa
{c}_{\rm A}'(\infty)&=& 0
\label{eq:boundary3}\\
{c}_{\rm A}(\infty) &=& 0. \label{eq:boundary4}
\end{eqnarray}
\end{subequations}

The three equations above, together with the three conditions of
Eqs.~(\ref{eq:h2}-\ref{eq:h4}), give us six conditions. The
Euler-Lagrange equations (Eq.~\ref{eq:euler2d}) are second order
differential equations for $c_{\rm B}(r)$ and $c_{\rm A}(r)$, which
have four constants. The solutions to these equations must be
integrated once to obtain $\eta(r)$, which yields six constants.
Finally, we determine the domain size $r_0$ by minimizing the total
free energy (after substituting the solution into the original free
energy) over $r_0$.  Hence we have a closed set of equations which
determine the height profiles $h_{\rm B}(r)$ and $h_{\rm A}(r)$ and
the impurity domain size $r_0$, for a given (arbitrary) frame area.

For the multi-domain problem the boundary conditions must be modified.
We still have Eqs~(\ref{eq:h2}-\ref{eq:h4}) above, but instead of
boundary conditions at $R=\infty$, we impose a symmetric profile at
the Wigner-Seitz cell boundaries:
\begin{equation}
c'_{\rm A}(R) = h'_{\rm A}(R)=0,\label{eq:2}
\end{equation}
replacing Eqs.~(\ref{eq:boundary3}-\ref{eq:boundary4}).
%%%%%%%%%%%%%%%%%%%%%%%%%%%%%%%%%%%%%%%%%%%%%%%%%%%%%%%%%%%%%%%%%%%%%%%%%%%%%
\section{Solutions}\label{app:b}
%%%%%%%%%%%%%%%%%%%%%%%%%%%%%%%%%%%%%%%%%%%%%%%%%%%%%%%%%%%%%%%%%%%%%%%%%%%%%
\subsection{Stripes}
%%%%%%%%%%%%%%%%%%%%%%%%%%%%%%%%%%%%%%%%%%%%%%%%%%%%%%%%%%%%%%%%%%%%%%%%%%%%%
For stripe domains the Wigner-Seitz cell is an impurity stripe of
width $2 x_0$ embedded inside a majority stripe with total with
$2X_0$.  The curvature is ${c}=d^2{\eta}/dx^2$, and the
Euler-Lagrange equations, Eqs.~(\ref{eq:bceuler1d}), are
\begin{subequations}
\label{eq:euler1d}
\begin{eqnarray}
 \kappa\frac{d^4{h}_{\rm A}}{d{x}^4} - {\sigma}
 \frac{d^2{h}_{\rm A}}{d{x}^2} &=&0, \label{eq:17}\\
 \kappa\frac{d^4{h}_{\rm B}}{d{x}^4} - ({\sigma}+{\tilde{\mu}})
 \frac{d^2{h}_{\rm B}}{d{x}^2} &=&0.\label{eq:18}
\end{eqnarray}
\end{subequations}
To calculate the free energy we only need the slope,
\begin{equation}
  \label{eq:eta}
  \eta = \frac{dh}{dx}.
\end{equation}
Eqs.~(\ref{eq:euler1d}) may be integrated once, with integration
constant zero (all odd derivatives $d^{(n)}h/dx^n$ vanish at the
origin). This leaves the following Euler-Lagrange equations:
\begin{subequations}
\label{eq:eulereta1d}
\begin{eqnarray}
  \frac{d^2{\eta}_{\rm A}}{d{\xi}^2} - \hat{\sigma}{\eta}_{\rm A}
  &=&0, \label{eq:15}\\ 
  \frac{d^2{\eta}_{\rm B}}{d{\xi}^2}-
  (\hat{\sigma}+\hat{\mu}){\eta}_{\rm B} &=&0,\label{eq:16}
\end{eqnarray}
\end{subequations}
where we have introduced the following dimensionless variables:
\begin{subequations}
\label{eq:dimensionless1}
  \begin{eqnarray}
    \xi &=& x c_0 \label{eq:12}\\
    \hat\sigma &=& \frac{\sigma}{\kappa c_0^2} \label{eq:13}\\
    \hat\mu &=& \frac{\tilde{\mu}}{\kappa c_0^2} \label{eq:14}\\
    \Xi &=& X_0 c_0.
  \end{eqnarray}
\end{subequations}
Note that $\tilde{\mu}$ and $\sigma$ both have dimensions of
energy/area, because we have implicitly assumed an incompressible
membrane with an unchanging area per molecule.  The solutions are
\begin{subequations}
\label{eq:sol1d}
\begin{eqnarray}
{\eta}_{\rm B} &=& \eta_0 \frac{\sinh {\zeta}_{\rm B}
\xi}{\sinh {\zeta}_{\rm B} {\xi}_0} \label{eq:10}\\
{\eta}_{\rm A}&=& - \eta_0 \frac{\sinh {\zeta}_{\rm A}
({\Xi}-{\xi})}{\sinh {\zeta}_{\rm A} ({\Xi}-{\xi}_0)},\label{eq:11}
\end{eqnarray}
\end{subequations}
where
\begin{eqnarray}
\eta_0 &=&  \left[{\zeta}_{\rm B} \coth{\zeta}_{\rm B} {\xi}_0
 + {\zeta}_{\rm A} \coth{\zeta}_{\rm A} ({\Xi}-{\xi}_0)\right]^{-1}
\label{eq:7}\\
{\zeta}_{\rm B}^2 &=& \hat\sigma+\hat\mu \label{eq:8}\\
{\zeta}_{\rm A}^2 &=& \hat\sigma.\label{eq:9}
\end{eqnarray}

The hyperbolic functions satisfy the conditions of zero slope at the
domain center $\xi=0$ and domain boundary $\xi=\Xi$ (Eq.~(\ref{eq:h2},
\ref{eq:2}), while the amplitude $\eta_0$ is determined by the
discontinuity of curvatures at $\xi=\xi_0$, according to
Eq.~(\ref{eq:boundary2}). 

After some rearrangement, the free energy
(Eq.~\ref{eq:ftot}) may be written as
\begin{equation}
\frac{G_0}{2 L \kappa c_0}  = \lambda + \mu\, {\xi}_0 -
\frac{\eta_0}{2}.
\end{equation}
The number of stripes is given by$ N = \frac{A_{\it fr}}{2 {\Xi} L}$ and
 the total free energy is $G=NG_0$. We can thus write the free energy
 per frame area $\hat{g}$, in dimensionless form, as
\begin{equation}
\hat{g}\equiv \frac{G}{\kappa c_0^2 A_{fr}}  =
  \frac{1}{\Xi}\left[\lambda + \mu\, {\xi}_0 - \frac{\eta_0}{2}\right].
\label{eq:fstripe}
\end{equation}
%%%%%%%%%%%%%%%%%%%%%%%%%%%%%%%%%%%%%%%%%%%%%%%%%%%%%%%%%%%%%%%%%%%%%%%%%%%%%
\subsection{Caplets}
%%%%%%%%%%%%%%%%%%%%%%%%%%%%%%%%%%%%%%%%%%%%%%%%%%%%%%%%%%%%%%%%%%%%%%%%%%%%%
We consider circular domains of inner radius $r_0$ and outer radius
$R$.  The Euler-Lagrange equations for the circular geometry,
Eqs.~(\ref{eq:euler2d}) are given 
by
\begin{equation}
\kappa({c}'_{\rm A} + r{c}''_{\rm A}) -\sigma r {c}_{\rm A} =0 ,
\label{eq:euler2db}
\end{equation}
and similarly for $c_{\rm B}$. In cylindrical coordinates, the mean
curvature is
\begin{equation}
{c}(r) =
\frac{1}{r} \frac{\partial}{\partial r}\left(r \frac{\partial
    h}{\partial r}\right). \label{eq:curvature}
\end{equation}
Equation~(\ref{eq:euler2db}) may be integrated once to give
\begin{equation}
  r\left\{\kappa c' - \sigma h'\right\}=\hbox{constant}.
\label{eq:2dintegral}
\end{equation}
Again, $h'(0)=c'(0)=0$ by symmetry, so the constant vanishes. As with
the stripe case, we only need the slope
\begin{equation}
  \eta = \frac{\partial h}{\partial r},
\end{equation}
in terms of which Eqs.~(\ref{eq:euler2d}) become
\begin{equation}
\frac{\partial^2\eta_{i}}{\partial\rho^2} +
\frac{1}{\rho}\frac{\partial\eta_{i}}{\partial\rho}
-\eta_{i}\left({\zeta}^2_{i} + \frac{1}{\rho^2}\right)  =0
\label{eq:euler2dc}
\end{equation}
for $i={\rm A},{\rm B}$, where $\rho=r c_0$. Implementing the boundary
conditions yields the following profiles:
\begin{subequations}
\begin{eqnarray}
\frac{\eta_{\rm B}}{\eta_0}&=&{\frac{I_1[{\zeta}_{\rm B}\rho]}{
I_1[{\zeta}_{\rm B}\rho_0]}}\\
\frac{\eta_{\rm A}}{\eta_0} &=&
\frac{K_1[{\zeta}_{\rm A}\rho]-A\, I_1[{\zeta}_{\rm A}\rho]}{
K_1[{\zeta}_{\rm A}\rho_0] - A\, I_1[{\zeta}_{\rm A}\rho_0]} ,
\end{eqnarray}
\label{eq:sol2d}
\end{subequations}
with
\begin{eqnarray}
  A &=& {\frac{K_1[{\zeta}_{\rm A}\bar{R}]}{
    I_1[{\zeta}_{\rm A}\bar{R}]}} \\
  \eta_0 &=& \frac{I_1[{\zeta}_{\rm B}\rho_0]}
    {{\zeta}_{\rm A} I_1[{\zeta}_{\rm B} \rho_0]
    \frac{K_0[{\zeta}_{\rm A}\rho_0]+ A\,
        I_0[{\zeta}_{\rm A}\rho_0]}{K_1
        \left[{\zeta}_{\rm A}\rho_0\right]-A\,I_1
        \left[{\zeta}_{\rm A}\rho_0\right]} +
      {\zeta}_{\rm B} I_0[{\zeta}_{\rm B}\rho_0]},
\end{eqnarray}
where $\bar{R}=Rc_0$ and $I_0, I_1, K_0,$ and $K_1$ are modified
Bessel functions \cite{AS}.  The free energy per caplet is given by
\begin{equation}
\frac{G_0}{2\pi\kappa} = \lambda \rho_0 +\frac12\rho_0
\left(\rho{\mu} - \eta_0\right).
\end{equation}
The number of caplets is $N= A_{\it fr} /(\pi \bar{R}^2)$, so the total
free energy is
\begin{equation}
  \hat{g} =\frac{G}{\kappa c_0^2A_{\it fr}} =
  \frac{\rho_0}{\bar{R}^2}\left[2\lambda +
    \rho_0 {\mu}-\eta_0\right].
\label{eq:fcap}
\end{equation}
Eqs.~(\ref{eq:fcap}, \ref{eq:fstripe}) are used in
Eqs.~(\ref{ELequations}) to evaluate the equilibrium domain sizes.
%%%%%%%%%%%%%%%%%%%%%%%%%%%%%%%%%%%%%%%%%%%%%%%%%%%%%%%%%%%%%%%%%%%%%%
%\bibliography{articles,books,membs,helfrich,slugsbibs} 
%\bibliographystyle{apsrev}
%%%%%%%%%%%%%%%%%%%%%%%%%%%%%%%%%%%%%%%%%%%%%%%%%%%%%%%%%%%%%%%%%%%%%%

%%%%%%%%%%%%%%%%%%%%%%%%%%%%%%%%%%%%%%%%%%%%%%%%%%%%%%%%%%%%%%%%%%%%%%
\end{document}